# Sliding induced multiple polarization states in two-dimensional ferroelectrics

Peng Meng[1,2,#], Yaze Wu[3,#], Renji Bian[1,#], Er Pan[1,#], Biao Dong[4], Xiaoxu Zhao[5], Jiangang Chen[1], Lishu Wu[6], Yuqi Sun[1], Qundong Fu[6], Qing Liu[1], Dong Shi[1], Qi Zhang[4], Yong-Wei Zhang[3,*], Zheng Liu[6,7,8,*], Fucai Liu[1,2,*]

1 School of Optoelectronic Science and Engineering, University of Electronic Science and Technology of China, Chengdu, China.

2 Yangtze Delta Region Institute (Huzhou), University of Electronic Science and Technology of China, Huzhou, China.

3 Institute of High Performance Computing, Agency for Science, Technology and Research (A*STAR), Singapore, Singapore.

4 School of Physics, Nanjing University, Nanjing, China.

5 School of Materials Science and Engineering, Peking University, Beijing, China.

6 School of Materials Science and Engineering, Nanyang Technological University, Singapore, Singapore.

7 CINTRA CNRS/NTU/THALES, UMI 3288, Research Techno Plaza, Singapore, Singapore.

8 Institute for Functional Intelligent Materials, National University of Singapore, Singapore, Singapore.

# These authors contributed equally: Peng Meng, Yaze Wu, Renji Bian, Er Pan.

*E-mail: zhangyw@ihpc.a-star.edu.sg, z.liu@ntu.edu.sg, fucailiu@uestc.edu.cn

**Abstract**

When the atomic layers in a non-centrosymmetric van der Waals structure slide against each other, the interfacial charge transfer results in a reversal of the structure's spontaneous polarization. This phenomenon is known as sliding ferroelectricity and it is markedly different from conventional ferroelectric switching mechanisms relying on ion displacement. Here, we present layer dependence as a new dimension to control sliding ferroelectricity. By fabricating 3R $MoS_2$ of various thicknesses into dual-gate field-effect transistors, we obtain anomalous intermediate polarization states in multilayer (more than bilayer) 3R $MoS_2$. Using results from *ab initio* density functional theory calculations, we propose a generalized model to describe the ferroelectric switching process in multilayer 3R $MoS_2$ and to explain the formation of these intermediate polarization states. This work reveals the critical roles layer number and interlayer dipole coupling play in sliding ferroelectricity and presents a new strategy for the design of novel sliding ferroelectric devices.

**Introduction**

Since the discovery of Rochelle salt in 1920, the study of ferroelectric materials has come a long way in its relatively short history of just one century[1-5]. The family of ferroelectric materials, whose spontaneous polarizations can be switched under electric fields, has since grown significantly to include perovskite oxides, hybrid perovskites, organic compounds, among many others[6,7]. These materials have shown tremendous industrial potential in applications such as non-volatile memory, actuators and negative capacitance field-effect transistors[8-14]. With an even shorter but equally successful history, two dimensional (2D) materials have also garnered extensive interest for their superior chemical and physical properties[15-20] since the first successful exfoliation of graphene in 2004[21]. As the intersect of ferroelectrics and 2D materials, 2D ferroelectrics have attracted considerable attention in the research community[4, 5, 22-24] in recent years. However, even though a lot of theoretical works have earlier predicted the existence of 2D ferroelectric materials, the experimental study of ultrathin layered ferroelectric material $CuInP_2S_6$ has only been conducted recently in 2016[13]. Since then, more 2D ferroelectric materials such as $MoTe_2$, $WTe_2$, $\alpha$-$In_2Se_3$, $\beta$-InSe, and $ReS_2$ have been reported[8-12, 25]. However, as the pre-requisite for ferroelectricity, i.e. non-centrosymmetric atomic structure, excludes a large proportion of the known materials, the number of 2D ferroelectric materials reported to date is woefully limited.

Recently, Li *et al.* predicted that specific stacking sequences of atomic layers in non-centrosymmetric van der Waals (vdW) materials lead to charge transfers between the atomic layers and give rise to out-of-plane spontaneous polarizations[26]. When the atomic layers slide against each other to reverse their stacking sequence, the charge transfer between them is also reversed. This flips the spontaneous polarization in the material and gives rise to "sliding ferroelectricity". Very recently, this concept was experimentally proved by several groups. Kenji *et al.* exfoliated a centrosymmetric AA' stacked h-BN crystal into monolayers and reassembled the monolayers to AB stacking using the tear-and-stack method[14]. As the crystal inversion symmetry was broken by the new AB stacking, the newly AB stacked h-BN showed distinct ferroelectric hysteresis under a sweeping electric field. This experimentally confirmed Li *et al.*'s predictions and proved sliding ferroelectricity's viability. Stern *et al.* also detected interfacial polarization in AB stacked h-BN using Kelvin probe force microscopy (KPFM) and observed that the sliding motion of ferroelectric domain wall could be driven by the Kelvin-probe under different biases[27]. The electrostatic potential variation between AB and BA stacked BN was also carefully measured by electrostatic force microscopy (EFM) by Woods *et al.*[28]. Since then, experimental observation of sliding ferroelectricity has been extended to semiconducting materials, such as 1T'-$ReS_2$[25] and rhombohedral stacking transition-metal dichalcogenides ($WSe_2$, $MoSe_2$, $WS_2$, $MoS_2$)[29]. The domain wall evolution in marginally twisted $MoS_2$ under transverse electric field was also investigated by back-scattered electron channelling contrast imaging (BSECCI)[30]. The abovementioned works demonstrated that by changing the stacking sequence of atomic layers, sliding ferroelectricity can be realized even from non-ferroelectric layered systems. On this account, the discovery of sliding ferroelectricity reveals a large collection of new 2D ferroelectrics and poses as a substantial advancement in addressing the present shortage in this class of functional materials. Furthermore, sliding ferroelectricity also introduces a new platform for investigating multiferroic phenomenon. For example, because the sliding motion of the atomic layers in few-layer $WTe_2$ is equivalent to an inversion operation to the energy band in **k**-space[31], the sign of its Berry curvature can be reversed and stored through sliding initiated ferroelectric switching. In another instance, selecting magnetic 2D materials for sliding ferroelectric applications can also couple these physical properties with sliding ferroelectricity, achieving multiferroics[26].

Nevertheless, despite the recent research efforts in sliding ferroelectricity, the fundamental question of layer dependence in sliding ferroelectricity is still open and should now be attended to. Here, we fabricate dual-gate field-effect transistors (FETs) from 3R $MoS_2$ flakes of different layer numbers and observe anomalous intermediate polarization states in multilayer (more than bilayer) 3R $MoS_2$. Subsequently, we conduct an extensive analysis of all possible interlayer-sliding-based ferroelectric switching pathways in trilayer 3R $MoS_2$ using *ab initio* density functional theory (DFT) calculations. From the results, a model is proposed to describe the ferroelectric switching mechanism in the trilayer system. This model is then generalized to describe ferroelectric switching in all multilayer 3R MoS$_2$ systems. The generalized model reveals the critical roles layer number and interlayer dipole coupling play in determining the properties of sliding ferroelectricity. The layer number is also identified as a good descriptor for the number of stable states and magnitudes of their spontaneous polarizations. The generalized layer-dependence in these properties paves the way for designing multistate ferroelectric devices in the future.

**Results and Discussion**

**Verification of atomic structure and piezoresponse in 3R MoS$_2$.** Various polymorphs of MoS$_2$ can be achieved by changing the stacking sequence of the MoS$_2$ monolayers, each of which is composed of a plane of molybdenum atoms sandwiched between two planes of sulphur atoms. In 2H phase MoS$_2$ (AA' stacking, space group *P63/mmc*), no out-of-plane spontaneous polarization is present (Fig. 1a, left panel) because inversion symmetry is preserved in even-layer systems and mirror symmetry is preserved in odd-layer systems[29]. In 3R phase MoS$_2$ (rhombohedral stacking, space group *R3m*), the lack of inversion and mirror symmetry allows for charge transfer between adjacent atomic layers, resulting in spontaneous polarization along the out-of-plane direction (Fig. 1a, right panel)[26]. For convenience, the relative position of each layer in a 3R stacked system is denoted as A, B and C, as shown in Fig. 1a. As illustrated, the position of each layer differs from its neighbour above by a translation of 1/3 unit cell in the $(\mathbf{b} - \mathbf{a})$ direction, where $\mathbf{a}$ and $\mathbf{b}$ are lattice vectors of the unit cell. By reversing the stacking order from cyclic (…AB<u>CABC</u>…) to anticyclic (…C<u>BACB</u>A…) through in-plane translation of atomic layers (see details in Fig. S1a and Fig. S1b), the spontaneous polarization of the system can be reversed. The 3R MoS$_2$ crystals are synthesized via the chemical vapor transport (CVT) method and their atomic structures are further verified using X-ray diffraction (XRD) (Fig. S2) and annular dark-field (ADF) scanning transmission electron microscopy (STEM) (Fig. 1b, detailed discussion in Fig. S3 and Supplementary

Note 1). To verify the presence of spontaneous polarization in 3R MoS$_2$, surface potential is measured using an atomic force microscope (AFM) operated in Kelvin probe mode (KPFM) (see Methods)[27]. As shown in Fig. 1d-f, the bilayer 3R flake with uniform thickness shows distinct domains with different surface potentials, corresponding to the up and down polarization domain ranges. Furthermore, we obtain Second Harmonic Generation (SHG) signal from flakes with different thicknesses (see Fig. 1c and Fig. S4), verifying the breaking of inversion symmetry in the samples[32]. The SHG intensities decrease with both the decrease in sample thickness and the increase in temperature, but remain observable even when temperature reaches 650 K as shown in Fig. 1c. This indicates that the ferroelectric transition temperature (T$_c$) for even the thinnest 3R MoS$_2$ (i.e. bilayer) is larger than 650 K, which is, to date, the highest in all known sliding ferroelectric materials to the best of our knowledge[25].

**Dual-gate FET performance of 3R MoS$_2$.** 3R MoS$_2$ is a semiconductor with a bandgap ranging from 1.1 eV to 1.6 eV depending on its thickness[33]. The Fermi level and carrier density of thin-film 3R MoS$_2$ can be easily tuned by the gate voltage. To investigate the intrinsic transport properties of 3R MoS$_2$, flakes with different layer numbers are fabricated into dual-gate FET devices as shown in Fig. 2a and b. The channel is connected by two graphite flakes, encapsulated by two h-BN dielectric layers of similar thickness and gated by two symmetric graphite gates. To regulate the vertical electric field (E$_\perp$) that penetrates the channel, the voltages on the symmetric gates are controlled by the following expression: $V_{bg} = AV_{tg} + B$. When $A = -1$ and $B = 0$, $V_{tg}$ and $V_{bg}$ are equal in amplitude but opposite in polarity ($V_{bg} = -V_{tg}$). By changing the gate voltages simultaneously, a sweeping vertical electric field is applied on the channel and the electrostatic doping on 3R MoS$_2$ is minimized. The electric field is defined by $E_\perp = \left(-V_{tg}/d_{tg} + V_{bg}/d_{bg}\right)/2$, where upward is defined as the positive electric field direction[9]. Furthermore, by changing the value of $B$, additional electrostatic doping can be introduced to investigate the influence of carrier density on the ferroelectric behaviour. Here, we note that in practice, $A$ usually deviates slightly from -1 to balance the voltage at the two gates in actual measurements because of the asymmetry in the gates caused by the following reasons. First, different h-BN flakes have inevitable small differences in thickness; Second, the top gate, unlike the bottom gate, is unable to regulate the Schottky barrier on the graphite/MoS$_2$ interface. The device parameters are summarized in Table S1 and the regular $I_d$-$V_d$ and $I_d$-$V_g$ results are shown in Fig. 2c and Fig. S5. For all the dual-gate measurements in this paper, the $E_\perp$ ramps down and then up in one loop. In Fig. 2d, all the 3R MoS$_2$ devices clearly exhibit butterfly-shaped hysteresis loops, which are absent in bilayer 2H MoS$_2$. This is evidence that the ferroelectricity in 3R MoS$_2$ originates from the rhombohedral stacking sequence, rather than the MoS$_2$/h-BN interface or the MoS$_2$/graphite Schottky barrier. Note that similar results have been reported in few-layer WTe$_2$[9]. In the sweeping loop, the stepwise drops in drain current represent the flipping of polarization, and the polarization flipping is realized by the movement of domain boundaries between domains with different polarizations[29]. In Fig. 2d, multiple stepwise drops can be observed in all devices. This implies that instead of the whole flake reversing stacking sequence abruptly, several domain boundaries move independently as the electric field increases. It is also evident that as the layer number increases, the device requires a stronger electric field to complete the flipping process across the whole flake. In the tetralayer device, the start and end points of the sweep measurement loop do not coincide because the electric field applied to the 3R MoS$_2$ is insufficient; any higher positive electric field in the tetralayer device will be beyond the insulating capability of h-BN. The presence of charge trapping, which can cause ferroelectric-like hysteresis loops, has also been excluded by changing the electric field ramping step size as shown in Fig. S6. Hysteresis loops caused by charge trapping usually shift considerably with the increase in charging time[34, 35]. However, no obvious shift is observed in our measurements after decreasing the ramping step size by one order of magnitude. The cycling performance for dual-gate 3R MoS$_2$ FET devices with different thickness are also examined and shown in Fig. S7a-c. The influence of electrostatic doping on the ferroelectricity in bilayer device is also investigated. As an n-type semiconductor, the $I_d$ in bilayer 3R MoS$_2$ increases by five orders of magnitude from $10^{-12}$ A to $10^{-7}$ A, when the doping bias (which is $B$ as mentioned above) changes from -0.3 V to 0.6 V (Fig. 2e and f). The carrier density is estimated to be $3.6\times10^{14}$ m$^{-2}$ when the bias is 0.6 V (detailed estimation method is discussed in Supplementary Note 2). To aid presentation, $I_d$ is normalized to the value at the start point of each loop for different values of bias (see Fig. 2g and h). The two primary flipping points marked by the white dotted line are located at -0.1 V nm$^{-1}$ (ramp down) and 0.19 V nm$^{-1}$ (ramp up), respectively. There are also some minor steps near the primary flipping point in the ramp-down process (below -0.1 V nm$^{-1}$). It can be observed that all the flipping points are almost fixed regardless of the changes in doping bias, suggesting that the coercive field of sliding ferroelectricity or the evolution of ferroelectric domain is independent of the variation in carrier density in bilayer 3R MoS$_2$ at room temperature.

**Anomalous intermediate states in multilayer 3R MoS$_2$.** Besides the dynamic behaviour under continuous sweeping electric field, the static behaviour is also investigated by applying a triangular electric field waveform as shown by the inset in Fig. 3a (an overall plot of the electric field is shown in Fig. S8). In the triangular waveform, the vertical electric field is switched on (on-field) and off (off-field) for the same pulse width in an alternating manner, while $I_d$ is monitored continuously. By averaging the $I_d$ over the on-field and off-field time periods respectively, the on-field and off-field $I_d$ trajectories against $E_\perp$ can be constructed (Fig. 3a-c). From the on-field plots (upper panels), the trend in the dynamic measurements in Fig. 2b is reproduced. This confirms the high repeatability and stability of the sliding ferroelectricity. Meanwhile, the off-field plots (lower panels) reveal an intrinsic relationship between the layer number and the sliding ferroelectricity.

In the bilayer device, $I_d$ shows an anti-clockwise rectangular hysteresis window with two different states, i.e. upward and downward polarizations. This is because there is only one MoS$_2$/MoS$_2$ interface in bilayer MoS$_2$; thus only one layer of interfacial dipoles can be formed and reversed under the electric field. Furthermore, the stepwise drops of $I_d$ in the off-field measurement coincide well with those in the on-field measurement. This implies that the movement of ferroelectric domain boundaries accompanied by the flipping of dipoles can affect the conductance of the channel. Similar switching dynamics has also been observed in twisted bilayer h-BN[14]. On the contrary, anomalous conductance states featuring non-monotonic changes in $I_d$ are found only in trilayer and tetralayer devices (as marked by the triangles in Fig. 3b and c). To verify that these observations are robust and not coincidental, we fabricated and tested one additional trilayer device. In this device, the abovementioned behaviour is also observed, as shown in Fig. S9. The presence of these anomalous conductance states is further confirmed in the dual-gate ferroelectric tunnelling junction (FTJ) device made from tetralayer 3R $MoS_2$ (see details in Fig. S10 and Supplementary Note 3). As in-plane domain boundary movement is present in all 3R $MoS_2$ systems but the anomalous conductance states are only formed in multilayer (thicker than bilayer) systems, we can conclude that domain boundary movement is not the root cause to the formation of anomalous conductance states. The deviation from the rectangular hysteresis window in trilayer and tetralayer 3R $MoS_2$ devices implies that new dipole configurations are formed at the anomalous conductance states. Furthermore, as $I_d$ does not vary monotonically with $E_\perp$ at the anomalous conductance states, these states should not be results of simple linear superposition of dipoles along the *c*-axis, but consequences of strong, unique coupling between the dipoles in the out-of-plane direction. Similar conclusions can be made when the stability of the states is tested as shown in Fig. 3d. During the test, the devices are regulated at four different electric fields (two start points, two different intermediate points and one end point in one loop). In each regulation, the electric field pulse lasts for 10s, then the corresponding retention is monitored for 1h. It is also evident here that the bilayer device has only two states, while both trilayer and tetralayer devices have multiple states, all of which are very stable. Based on these experimental observations and analyses above, we propose an anti-parallel polarization model to describe the sliding ferroelectric switching process. Taking trilayer 3R $MoS_2$ as an example, there exist two $MoS_2/MoS_2$ interfaces between the three atomic layers, hence two dipoles are formed along the *c*-axis in the initial state as shown in Fig. 3e. As the vertical electric field decreases, instead of the polarization of the entire system flipping collectively, the different atomic layers slide sequentially and the dipoles between the layers are flipped one interface at a time. The anti-parallel arrangement of dipoles along the *c*-axis gives rise to an intermediate polarization state (as shown in Fig. 3e) that leads to the formation of the anomalous conductance states observed in the ramp-down process in Fig. 3b-c. The same mechanism can be applied to the ramp-up process. This model can also be generalized to describe ferroelectric switching in thicker (more than trilayer) 3R MoS$_2$.

**Theoretical analysis on ferroelectric switching pathways in trilayer 3R MoS$_2$.** To justify the feasibility of the anti-parallel polarization model, we use the trilayer 3R MoS$_2$ as a starting point and analyse the different types of interlayer translations available, before proposing a ferroelectric switching pathway that realizes the model's key characteristics. Density functional theory calculations are performed to provide a quantitative description of the structural, thermodynamic, dynamic and polarization changes in the ferroelectric switching process. The trilayer-specific model is then generalized to describe the ferroelectric switching process in $n$-layer 3R MoS$_2$. Because the intermediate polarization states are associated with strong coupling between the out-of-plane dipoles, the theoretical analysis focuses on ferroelectric switching through the layer sliding mechanism. The influence of domain boundaries is hence neglected. A detailed discussion on the origin of spontaneous polarization in 3R MoS$_2$ is also available in Supplementary Note 4. In trilayer 3R MoS$_2$, ferroelectric switching by interlayer sliding can be achieved by three types of relative atomic layer translations, as presented in the insets of Fig. 4a-c. In the first type (Type I), two adjacent atomic layers remain stationary while one atomic layer translates, leading to 1 sliding interface (see Fig. 4a inset). In the second type (Type II), two non-adjacent atomic layers translate simultaneously in opposite directions while the other atomic layer remains stationary, leading to 2 sliding interfaces occurring at the same time (see Fig. 4b inset). In the third type (Type III), two adjacent atomic layers translate simultaneously in opposite directions while the other atomic layer remains stationary, also leading to 2 sliding interfaces at the same time (see Fig. 4c inset). The potential energy surface (PES) experienced by each moving atomic layer in each of these translation types is calculated (see Fig. 4a-c). From the PES, it can be observed that the translations of atomic layers along the $\pm(\mathbf{b} - \mathbf{a})$ direction (as marked by the white and red arrows) exhibit the lowest energy barriers for all the cases. This suggests that regardless of translation type, the atomic layers should translate along $\pm(\mathbf{b} - \mathbf{a})$ as it is the most energetically favourable direction. Details of the PES calculation are presented in Supplementary Note 5. Furthermore, the PES experienced by the translating atomic layer in Type I translation is very similar to that in a bilayer 3R MoS$_2$, where there is also only one sliding interface[36]. This suggests that when there is only one sliding interface, the presence of additional atomic layers does not significantly affect the PES experienced by the moving atomic layer. Therefore, Type I translation can be generalized to describe the ferroelectric switching process in 3R MoS$_2$ with any number of atomic layers. This is especially important because Type I translation fits our proposed model and presents the lowest energy barrier among the three translation types.

To demonstrate the details of the three types of translations, we present one ferroelectric switching pathway for each translation type. In Path 1 (see Fig. 4d), the ramp-up and ramp-down processes are each divided into two steps, where each step comprises a Type I translation. Take the ramp-up process for example; in the first step, the bottom layer is translated by $\boldsymbol{\delta} = \frac{1}{3}(\mathbf{b} - \mathbf{a})$. This moves the bottom layer from A to C and switches the stacking order between the bottom and middle atomic layers from the cyclic AB to the anticyclic CB, resulting in the reversal of spontaneous polarization at the interface between the two layers from downward to upward. Here, we note the formation of an intermediate polarization state (CBC) that possesses two oppositely aligned interlayer dipoles along the *c*-axis as shown in Fig. 4d. In the second step, the top layer is translated by $-\boldsymbol{\delta}$ from C to A. The stacking order between the middle and top layer is hence reversed from the cyclic BC to anticyclic BA. The serial combination of the two steps completes the stacking reversal of the cyclically stacked initial state (ABC) to the anticyclically stacked final state (CBA) and hence reverses the spontaneous polarization in the entire trilayer from downward to upward. Here, for completeness, we also introduce the formation of the non-equivalent intermediate polarization state (ABA) in the ramp-down process. The formation of intermediate polarization states aligns well with the experimental observations and provides an opportunity to study the intermediate polarization states in detail. Path 2 (see Fig. 4d) demonstrates a Type II translation where the non-adjacent top and bottom layers are concurrently translated by $-\boldsymbol{\delta}$ and $\boldsymbol{\delta}$ respectively. This reverses the cyclic stacking order (ABC) in the initial state to the anticyclic stacking order (CBA) in one step. Path 3 (see Fig. 4d) demonstrates a Type III translation where the adjacent top and middle layers are translated concurrently by $\boldsymbol{\delta}$ and $-\boldsymbol{\delta},$ respectively. This reverses the cyclic stacking order (ABC) to the anticyclic stacking order (ACB) in one step, reversing the spontaneous polarization of the trilayer. An equivalent variation to Path 3 is shown in Fig. S11. As seen in Fig. 4e, the three paths show significantly different energy profiles, with Path 1 being the minimum energy path (MEP). In Path 1, each of the intermediate polarization states (CBC and ABA) reside in a valley between two barriers with the same height (15.4 meV). The total energy of the intermediate polarization state is only 0.2 meV higher than those of the initial and final states. On the other hand, Path 2 and 3 present energy barriers of 30.6 meV and 83.3 meV, respectively. This confirms the kinetic favourability of Path 1 and hence validates the feasibility of our model Fig. 3e. By residing in valleys between energy barriers and having similar energy to the initial and final states, the intermediate polarization states in Path 1 also demonstrate good thermodynamic stability that allows for their extended existence, which agrees well with the excellent stability of the anomalous conduction states shown in Fig. 3d.

To understand the superior thermodynamic performance of Path 1 compared to Path 2 and 3, we relate the energy profile of each path to the electronic interactions between the atomic layers. As discussed in detail in Supplementary Note 6, the energy barriers generated during relative layer movements originate from the short-range electronic repulsion between the adjacent layers. Therefore, the repulsion is limited to the sliding interface and is independent from the movements or stacking orders of other layers. The sliding interfaces are marked by the grey horizontal lines in Fig. 4d. In Path 3, the high energy barrier is attributed to the strong electronic repulsion associated with the AA stacking between the top and middle atomic layers formed during the switching process. On the other hand, in Path 1 and 2 where AA stacking is absent, Path 1 is observed to have only one sliding interface at any point in the ferroelectric switching process (Fig. 4d), while Path 2 has two such sliding interfaces. As the electronic repulsions between the adjacent layers are short ranged and independent from other layers, the energy barrier generated at each sliding interface is additive. This is confirmed by the fact that Path 2 has twice the energy barrier height as Path 1. In summary, the kinetic favourability of Path 1 can be attributed to the absence of AA stacking and the minimization of sliding interfaces.

**Generalized model of ferroelectric switching process in $n$-layer 3R MoS$_2$.** With these insights, Path 1 can be generalized to describe the ferroelectric switching mechanism in all multilayer 3R MoS$_2$. In this generalized model, AA stacking between adjacent atomic layers is actively avoided and the number of sliding interfaces is kept to one to minimize the height of the energy barriers generated. As such, the switching process of an $n$-layer system is divided into $(n-1)$ steps. In each step, the $n$layer 3R MoS$_2$ is divided into one moving block and one stationary block. The moving block is translated along $\boldsymbol{\delta}$ or $-\boldsymbol{\delta}$ to reverse the cyclic/anticyclic layer stacking order at the sliding interface between the blocks. A stable intermediate polarization state is formed at the end of the step. In the next step, new moving and stationary blocks are defined so another round of translation takes place. As the interlayer repulsion at each sliding interface is independent from each other, there is no constraint on the location of the subsequent sliding interface after each step, as long as the spontaneous polarization of the system continues to be switched towards the desired direction. The ferroelectric switching process is completed when the stacking order of the entire system is reversed. Applying this generalized model to tetralayer ($n = 4$) and pentalayer ($n' = 5$) 3R MoS$_2$ leads to the formation of three ($n - 1 = 3$) and four ($n' - 1 = 4$) energy barriers, respectively, as shown in Fig. 4e. Demonstration of this generalized model in trilayer, tetralayer and pentalayer 3R MoS$_2$ is illustrated in Fig. S12. The presence of the $n$-layer systems' $n$ stable spontaneous polarizations and examples of their associated atomic structures present an explanation to the formation of multiple polarization states observed experimentally in tetralayer 3R MoS$_2$ (see Fig. 3c). Compared to the bilayer limit of 15.0 meV cell$^{-1}$ in ferroelectric switching barrier height, only a small increase of less than 1 meV cell$^{-1}$ in switching barrier height is observed in these thicker systems (see Fig. S13). This insensitivity of the barrier heights to the number of atomic layers can be attributed to the independence of the electronic repulsion at each sliding interface from stacking orders between other layers.

One key highlight of our generalized model is the formation of multiple intermediate polarization states with varied spontaneous polarizations. As seen in Fig. 4f, the intermediate polarization states in Fig. 4e show different spontaneous polarizations depending on the number of ferroelectric switching steps that precedes it. By residing in the valleys between energy barriers (see Fig. 4e, hollow diamond symbols), these intermediate polarization states are thermodynamically stable and hence are able to preserve their spontaneous polarizations. As seen in Fig. 4e and f, there are $n$ stable spontaneous polarizations for a $n$-layer 3R MoS$_2$. For each spontaneous polarization ($P_z$) in a $n$-layer system (see Fig. 4g), the number of non-equivalent polarization states (such as CBC and ABA in Path 1, Fig. 4d) satisfies the condition $N_P = \mathbb{C}_k^{n-1}$, where $k$ is the number of layers in cyclic stacking order and $\mathbb{C}_k^n = \frac{n!}{k!(n-k)!}$. The total number of possible polarization states ($N$) in the entire ferroelectric switching process is therefore $\sum_{k=n-1}^{0} \mathbb{C}_k^{n-1} = 2^{n-1}$. The exponential scaling of total number of polarization states with number of layers, the states' thermodynamic stability, and their varied magnitudes of spontaneous polarization provide a blueprint for the sliding ferroelectric device developments in the future.

In summary, we report robust sliding ferroelectricity in dual-gate FET devices based on 3R MoS$_2$. Besides high stability and good retention, the 3R MoS$_2$ in these devices exhibits high T$_c$ (beyond 650 K) and strong immunity to the variation in carrier density. In the ferroelectric switching process, anomalous intermediate polarization states and strong coupling among interlayer dipoles are discovered in multilayer (more than bilayer) 3R MoS$_2$. Using results from density functional theory, we propose a generalized model to describe the ferroelectric switching process in multilayer 3R MoS$_2$. Using this model, we relate the number of stable states and spontaneous polarizations generated during the ferroelectric switching process to the number of layers. By analysing the switching behaviour in different layer 3R MoS$_2$, the layer number of the material and the coupling of interlayer dipoles are suggested as new dimensions to control sliding ferroelectricity.

Note: During the review process of this manuscript, we become aware of a similar result[37] published in Nature.

**Methods**

**Device fabrication:** The 3R MoS$_2$ crystals were grown by CVT method, the h-BN crystals were purchased from HQ Graphene, and the graphite crystals were purchased from NGS company. The 3R MoS$_2$, h-BN and graphite crystals were exfoliated onto SiO$_2$ (285 nm)/Si substrates. The layer number of 3R MoS$_2$ was identified by optical contrast. The thicknesses of h-BN flakes were measured by atomic force microscopy in tapping mode. After all materials were prepared, the flakes were picked up layer by layer with a poly(bisphenol A carbonate) (PC)-film-covered polydimethylsiloxane (PDMS) stamp on a glass slide at 90 °C. The whole stack was released on a SiO$_2$ (285 nm)/Si (p+ doped) substrate at 180 °C followed by rinsing in chloroform. The Cr/Au (8 nm/50 nm) electrodes were defined on the stack using digital micro-mirror device (DMD) lithography followed by the metal e-beam evaporation and lift-off process. Finally, the device was annealed in a mixed Ar/H$_2$ (9:1) atmosphere at 100 sccm and 300 °C for 6h to reduce the bubbles.

**SHG Measurement:** SHG signals were measured by HAMAMATSU H12386-110 photomultiplier tube or a spectrometer (Princeton Instrument HRS-750-MS) with a X50 objective lens (NA=0.5) at temperatures ranging from 298 K to 650 K. The Ti:sapphire laser (wavelength: 800 nm, repetition rate 80 MHz) with horizontal polarization was adopted. The 3R MoS$_2$ crystal was mechanically exfoliated on SiO$_2$(285 nm)/Si substrates. The layer number of the sample was identified by optical contrast.

**KPFM measurements:** The surface potential was measured using an atomic force microscope (Asylum MFD-3D Origin) under the Kelvin Probe Force mode (KPFM). We employ the HQ:NSC14/Cr-Au conductive probe, which has a force constant of 5 N/m and a resonance frequency of~160 kHz. The KPFM adds an additional feedback loop compared to the normal AC mode to record the change in the surface potential of the sample. The first pass is similar to the standard

AC mode, recording topography. In the second pass, the tip is raised to a fixed height from the sample surface to record the potential difference between the tip of the probe and the sample.

**Transport measurements:** The electric transport properties were measured with FS-Pro 380 semiconductor parameter analyser in a vacuum chamber of $10^{-2}$ Torr. The triangular electric field waveform was composed by on-field and off-field sweep processes, whose segment periods were 5 sec and sampling time was 0.1 sec. When calculating the average drain current of a segment, the first 10 data points were dropped to eliminate the influence of the charging/discharging of the system. The drain voltage for dual-gate FET devices and dual-gate FTJ device were 10 mV and 1 mV, respectively.

**XRD:** X-ray Diffraction was carried out on an X-ray diffractometer (DX-27) at 30 mA and 40 kV using monochromatic Cu Kα radiation at a scanning rate of 0.03° sec$^{-1}$ from 10° to 80°. The 3R MoS$_2$ crystal was ground to powder before measurements.

**STEM:** The sample of 3R MoS$_2$ was made by drop casting. The atomic-resolution ADF-STEM imaging was performed on an aberration-corrected ARM200F, equipped with a cold field-emission gun operating at 80 kV.

**Simulation:** The density functional theory calculations were performed with the plane-wave pseudopotential code Vienna Ab initio Simulation Package (VASP)[38-41], employing the Perdew-Burke-Ernzerhof (PBE) generalized gradient approximation (GGA)[42] for the exchange-correlation functional. Van der Waals correction was incorporated by the Grimme's D3 method[43]. We used an energy cut-off of 520 eV and a Monkhorst-Pack **k**-point mesh of 12×12×1 for all calculations. Relaxation of the atomic structures at the endpoints of each ferroelectric switching process was performed with a force convergence criterion of 0.001 eV Å$^{-1}$. The geometries along the ferroelectric switching pathways were calculated using the climbing nudged elastic band (NEB) method[44] with a force convergence criterion of 0.01 eV Å$^{-1}$. A criterion of $10^{-10}$ eV was used for the convergence of the self-consistent cycles. The calculations were performed with vertical vacuum separation of about 20 Å between the 2D materials. The Berry phase method was used to evaluate the crystalline spontaneous polarization[45]. The spatial charge distribution of the systems was calculated using the Bader Charge analysis[46].

## Data Availability Statement

The data that support the findings of this study are available from the corresponding authors upon reasonable request.

## References

1. Martin LW, Rappe AM. Thin-film ferroelectric materials and their applications. *Nature Reviews Materials* **2**, (2016).
2. Setter N, *et al.* Ferroelectric thin films: Review of materials, properties, and applications. *Journal of Applied Physics* **100**, (2006).
3. Bowen CR, Kim HA, Weaver PM, Dunn S. Piezoelectric and ferroelectric materials and structures for energy harvesting applications. *Energy Environ Sci* **7**, 25-44 (2014).
4. Guan Z, *et al.* Recent Progress in Two-Dimensional Ferroelectric Materials. *Advanced Electronic Materials* **6**, 1900818 (2019).
5. Xue F, He J-H, Zhang X. Emerging van der Waals ferroelectrics: Unique properties and novel devices. *Applied Physics Reviews* **8**, 021316 (2021).
6. Martin LW, Rappe AM. Thin-film ferroelectric materials and their applications. *Nature Reviews Materials* **2**, 16087 (2016).
7. Horiuchi S, Tokura Y. Organic ferroelectrics. *Nature Materials* **7**, 357-366 (2008).
8. Yuan S, *et al.* Room-temperature ferroelectricity in $MoTe_2$ down to the atomic monolayer limit. *Nature Communications* **10**, 1775 (2019).
9. Fei Z, *et al.* Ferroelectric switching of a two-dimensional metal. *Nature* **560**, 336-339 (2018).
10. Zhou Y, *et al.* Out-of-Plane Piezoelectricity and Ferroelectricity in Layered alpha-$In_2Se_3$ Nanoflakes. *Nano letters* **17**, 5508-5513 (2017).
11. Zheng C, *et al.* Room temperature in-plane ferroelectricity in van der Waals $In_2Se_3$. *Science advances* **4**, eaar7720 (2018).
12. Hu H, Sun Y, Chai M, Xie D, Ma J, Zhu H. Room-temperature out-of-plane and in-plane ferroelectricity of two-dimensional β-InSe nanoflakes. *Applied Physics Letters* **114**, 252903 (2019).
13. Liu F, *et al.* Room-temperature ferroelectricity in $CuInP_2S_6$ ultrathin flakes. *Nature communications* **7**, 12357 (2016).
14. Yasuda K, Wang X, Watanabe K, Taniguchi T, Jarillo-Herrero P. Stacking-engineered ferroelectricity in bilayer boron nitride. *Science* **372**, 1458-1462 (2021).
15. Novoselov KS, Mishchenko A, Carvalho A, Castro Neto AH. 2D materials and van der Waals heterostructures. *Science* **353**, aac9439 (2016).
16. Li L, *et al.* Black phosphorus field-effect transistors. *Nature nanotechnology* **9**, 372-377 (2014).
17. Wang QH, Kalantar-Zadeh K, Kis A, Coleman JN, Strano MS. Electronics and optoelectronics of two-dimensional transition metal dichalcogenides. *Nat Nanotechnol* **7**, 699-712 (2012).

18. Geim AK, Novoselov KS. The rise of graphene. *Nature Materials* **6**, 183-191 (2007).
19. Radisavljevic B, Radenovic A, Brivio J, Giacometti V, Kis A. Single-layer MoS$_2$ transistors. *Nature nanotechnology* **6**, 147-150 (2011).
20. Geim AK, Grigorieva IVJN. Van der Waals heterostructures. **499**, 419-425 (2013).
21. Novoselov KS, *et al.* Electric field effect in atomically thin carbon films. *Science* **306**, 666-669 (2004).
22. Kim JY, Choi M-J, Jang HW. Ferroelectric field effect transistors: Progress and perspective. *APL Materials* **9**, 021102 (2021).
23. Qi L, Ruan S, Zeng YJ. Review on Recent Developments in 2D Ferroelectrics: Theories and Applications. *Advanced Materials* **33**, 2005098 (2021).
24. Wu M. Two-Dimensional van der Waals Ferroelectrics: Scientific and Technological Opportunities. *ACS Nano* **15**, 9229-9237 (2021).
25. Wan Y, *et al.* Room-Temperature Ferroelectricity in 1T'-ReS$_2$ Multilayers. *Physical Review Letters* **128**, 067601 (2022).
26. Li L, Wu M. Binary Compound Bilayer and Multilayer with Vertical Polarizations: Two-Dimensional Ferroelectrics, Multiferroics, and Nanogenerators. *ACS Nano* **11**, 6382-6388 (2017).
27. Vizner Stern M, *et al.* Interfacial ferroelectricity by van der Waals sliding. *Science* **372**, 1462-1466 (2021).
28. Woods CR, *et al.* Charge-polarized interfacial superlattices in marginally twisted hexagonal boron nitride. *Nature Communications* **12**, 347 (2021).
29. Wang X, *et al.* Interfacial ferroelectricity in rhombohedral-stacked bilayer transition metal dichalcogenides. *Nature nanotechnology*, 1-5 (2022).
30. Weston A, *et al.* Interfacial ferroelectricity in marginally twisted 2D semiconductors. *Nature nanotechnology*, 1-6 (2022).
31. Xiao J, *et al.* Berry curvature memory through electrically driven stacking transitions. *Nature Physics* **16**, 1028-1034 (2020).
32. Shi J, *et al.* 3R MoS$_2$ with Broken Inversion Symmetry: A Promising Ultrathin Nonlinear Optical Device. *Advanced Materials* **29**, 1701486 (2017).
33. Rahman IA, Purqon A. First principles study of molybdenum disulfide electronic structure. *J Phys Conf Ser* **877**, 012026 (2017).
34. Alam MNK, *et al.* On the Characterization and Separation of Trapping and Ferroelectric Behavior in HfZrO FET. *IEEE Journal of the Electron Devices Society* **7**, 855-862 (2019).
35. Warren WL, *et al.* Voltage shifts and defect-dipoles in ferroelectric capacitors. *MRS Online Proceedings Library (OPL)* **433**, (1996).
36. Levita G, Cavaleiro A, Molinari E, Polcar T, Righi MC. Sliding Properties of MoS$_2$ Layers: Load and Interlayer Orientation Effects. *The Journal of Physical Chemistry C* **118**, 13809-13816 (2014).
37. Deb S, *et al.* Cumulative polarization in conductive interfacial ferroelectrics. *Nature*, (2022).
38. Kresse G, Hafner J. Ab initio molecular dynamics for liquid metals. *Physical Review B* **47**, 558-561 (1993).
39. Kresse G, Furthmüller J. Efficiency of ab-initio total energy calculations for metals and semiconductors using a plane-wave basis set. *Computational Materials Science* **6**, 15-50 (1996).
40. Kresse G, Furthmüller J. Efficient iterative schemes for ab initio total-energy calculations using a plane-wave basis set. *Physical Review B* **54**, 11169-11186 (1996).
41. Kresse G, Joubert D. From ultrasoft pseudopotentials to the projector augmented-wave method. *Physical Review B* **59**, 1758-1775 (1999).
42. Perdew JP, Burke K, Ernzerhof M. Generalized gradient approximation made simple. *Physical review letters* **77**, 3865 (1996).

43. Grimme S. Semiempirical GGA-type density functional constructed with a long-range dispersion correction. *Journal of Computational Chemistry* **27**, 1787-1799 (2006).
44. Henkelman G, Uberuaga BP, Jónsson H. A climbing image nudged elastic band method for finding saddle points and minimum energy paths. *Journal of Chemical Physics* **113**, 9901-9904 (2000).
45. King-Smith RD, Vanderbilt D. Theory of polarization of crystalline solids. *Physical Review B* **47**, 1651 (1993).
46. Tang W, Sanville E, Henkelman G. A grid-based Bader analysis algorithm without lattice bias. *Journal of Physics: Condensed Matter* **21**, 084204 (2009).

**Acknowledgements**

This work was supported by the National Natural Science Foundation of China (12161141015, 62074025) and the National Key Research & Development Program (2021YFE0194200, 2020YFA0309200), the Applied Basic Research Program of Sichuan Province (2021JDGD0026), the Postdoctoral Innovative Talent Supporting Program (BX20190060) and Sichuan Province Key Laboratory of Display Science and Technology. This work was also partially supported by the National Research Foundation, Singapore under Award No. NRF-CRP24-2020-0002, and Singapore A*STAR SERC CRF Award. The use of computing resources at the National Supercomputing Centre Singapore is gratefully acknowledged. Z.L. acknowledges the support from National Research Foundation, Singapore, under its Competitive Research Programme (CRP) (NRF-CRP22-2019-0007, NRF-CRP22-2019-0004). This research is also supported by A*STAR under its AME IRG Grant (Project No. A2083c0052), and the Ministry of Education, Singapore, under its Research Centre of Excellence award to the Institute for Functional Intelligent Materials. Project No. EDUNC-33-18-279-V12.

**Author contributions**

Z.L., F.L. and Y.Z. supervised the project. F.L. and P.M. conceived the idea and designed the experiments. P.M. fabricated the dual-gate FET devices, performed the electronic transport measurements and analysed the data. Y.W. performed the DFT calculations and theoretical analysis. Y.Z. provided guidance on the theoretical analysis and calculations. R.B. and Q.L. synthesized the 3R MoS$_2$ crystal, conducted the material characterization with the help of Q.F. and L.W. R.B. fabricated the dual-gate FTJ devices and carried out the tunnelling measurement with the help of J.C. E.P. performed the PFM measurements with the help of Y.S. and D.S. B.D. and Q.Z. performed the SHG measurements. X.Z. performed the STEM measurements. P.M., Y.W. and F.L. wrote the manuscript with the input from all authors. All authors discussed the results.

**Competing interests**

The authors declare no competing interests.

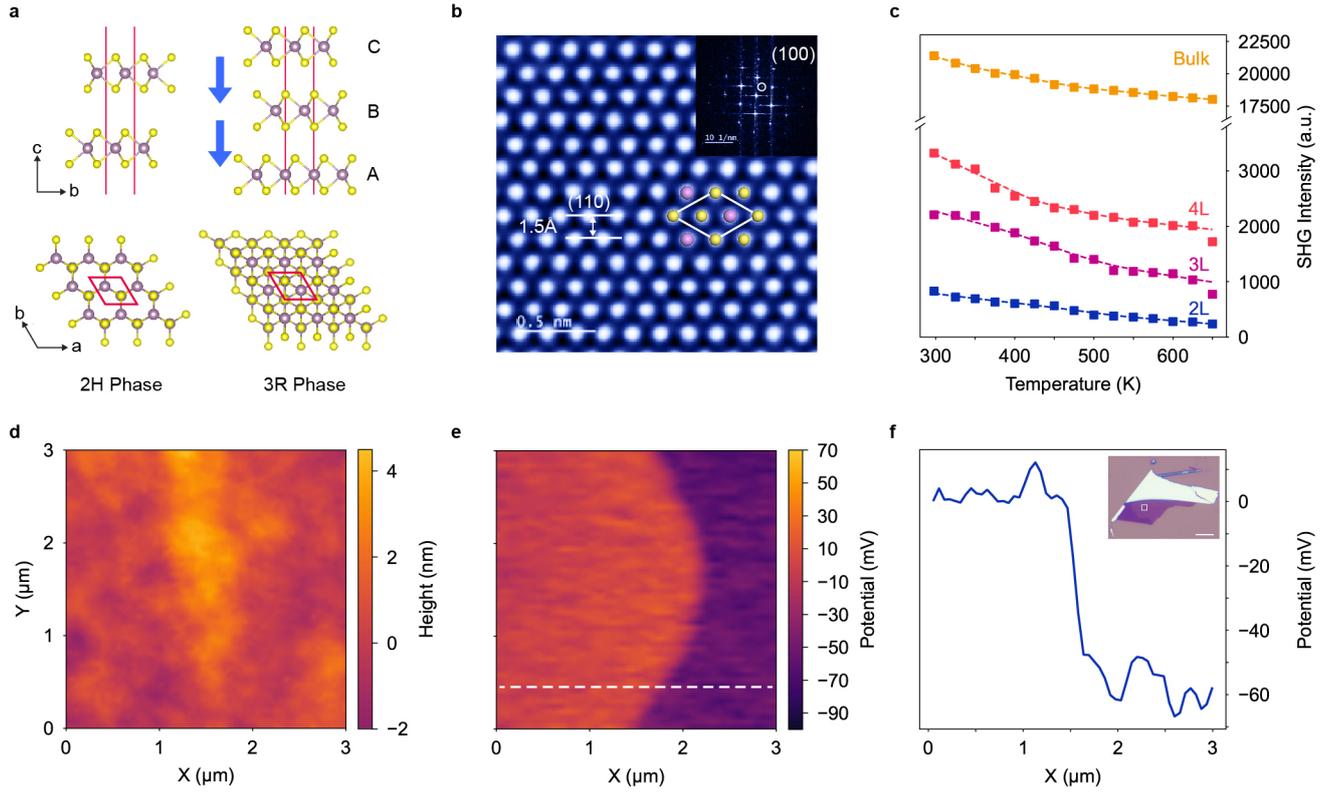

**Fig. 1. Verification of ferroelectricity in 3R MoS2. (a)** Top and side view schematic of 2H and 3R MoS2 atomic structures. The yellow and purple spheres represent the sulphur and molybdenum atoms, respectively. The unit cells are marked by the red diamond boxes. The blue arrows indicate the direction of spontaneous polarization. **(b)** Atomic-resolution ADF-STEM image of thin 3R MoS2 flake. The unit cell is marked by the white diamond box, the yellow and purple spheres represent sulphur and molybdenum atoms, respectively. The d-spacing of the (110) plane is 1.5 Å. The inset shows the FFT patterns, where the (100) plane is marked. **(c)** SHG intensity as a function of temperature from 298 K to 650 K in 3R MoS2 of different layer numbers. **(d)** AFM topology and **(e)** KPFM amplitude of bilayer 3R $MoS_2$ flake. **(f)** Surface potential profile along the white dotted line marked in (e). The microscope image of the bilayer flake is shown as the inset in (f). The white box in the image indicates the KPFM scan area, the scar bar is 10μm.

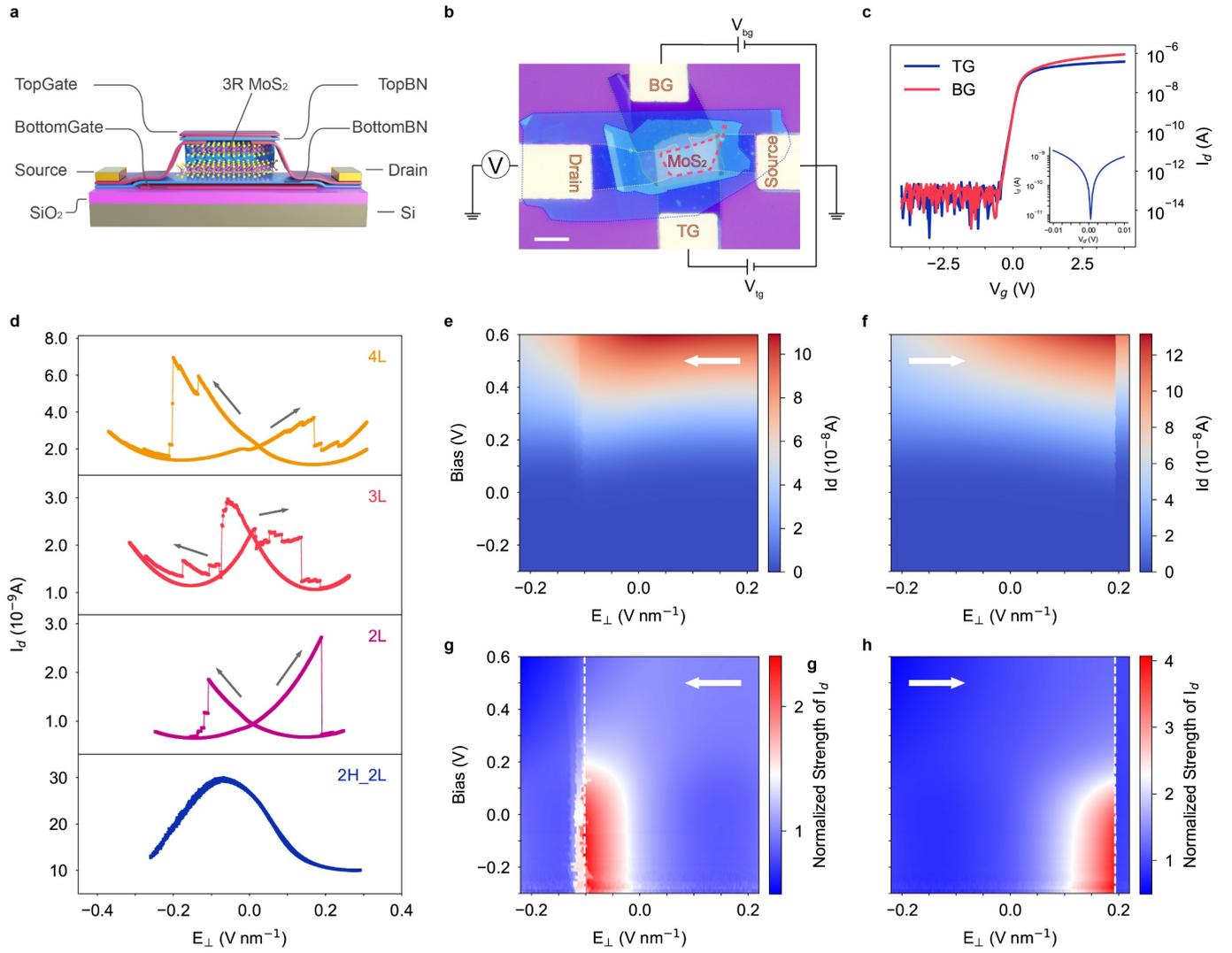

**Fig. 2. Dynamic transport properties of dual-gate FET devices. (a)** Schematic of a typical 3R MoS$_2$ dual-gate device. **(b)** Micrograph image of a bilayer device. The scale bar is 10 μm. **(c)** Drain current (I$_d$) as a function of top gate and bottom gate voltages in bilayer 3R MoS$_2$. The inset is the plot of I$_d$ against V$_d$ in bilayer 3R MoS$_2$ without gating voltage. **(d)** Drain current (I$_d$) as a function of vertical electric field (E$_\perp$) in bilayer 2H MoS$_2$ and 3R MoS$_2$ of different layer numbers in a loop of sweep measurement. Upward is defined as the positive electric field direction. Drain current (I$_d$) as a function of vertical electric field (E$_\perp$) and doping bias (Bias) in **(e)** forward sweep and **(f)** backward sweep for bilayer device. For better illustration, normalized strength of drain current is plotted as a function of vertical electric field (E$_\perp$) and doping bias (Bias) in **(g)** forward sweep and **(h)** backward sweep for bilayer device. The dashed lines mark the polarization flipping points. The drain current is normalized to the drain current value at the start point of each sweep.

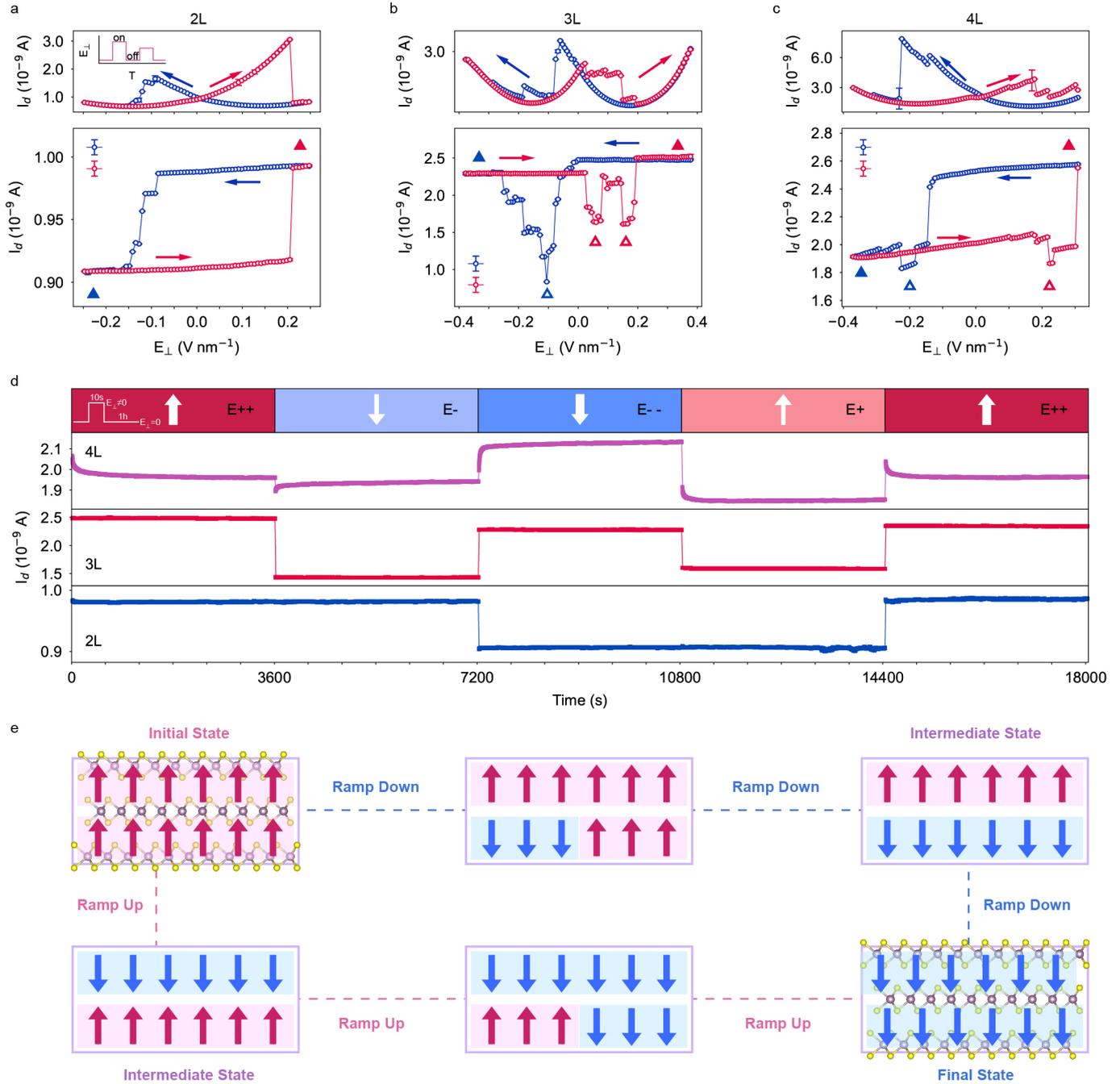

**Fig. 3. Static transport properties of dual-gate FET devices in the ferroelectric switching process. (a)-(c)** The averaged $I_d$ as a function of $E_\perp$ in on-field (upper panel) measurements and off-field (lower panel) measurements in devices with different layer numbers. The solid triangles indicate the locations of the initial and final states in the loop, and the hollow triangles indicate the locations of the intermediate anomalous states. The arrows indicate the sweep direction. The inset in (**a**) is a schematic of the triangular electric field waveform applied on the devices. (**d**) Retention of different states in different layer number dual-gate FETs for 1h. The inset in (**d**) illustrates an electric field with a pulse width of 10s followed by retention monitoring of 1h. In the plot of $I_d$ against time, data for the 10s pulse is dropped

for better illustration. **(e)** The model of ferroelectric switching and the evolution of dipole arrangement in trilayer 3R MoS$_2$.

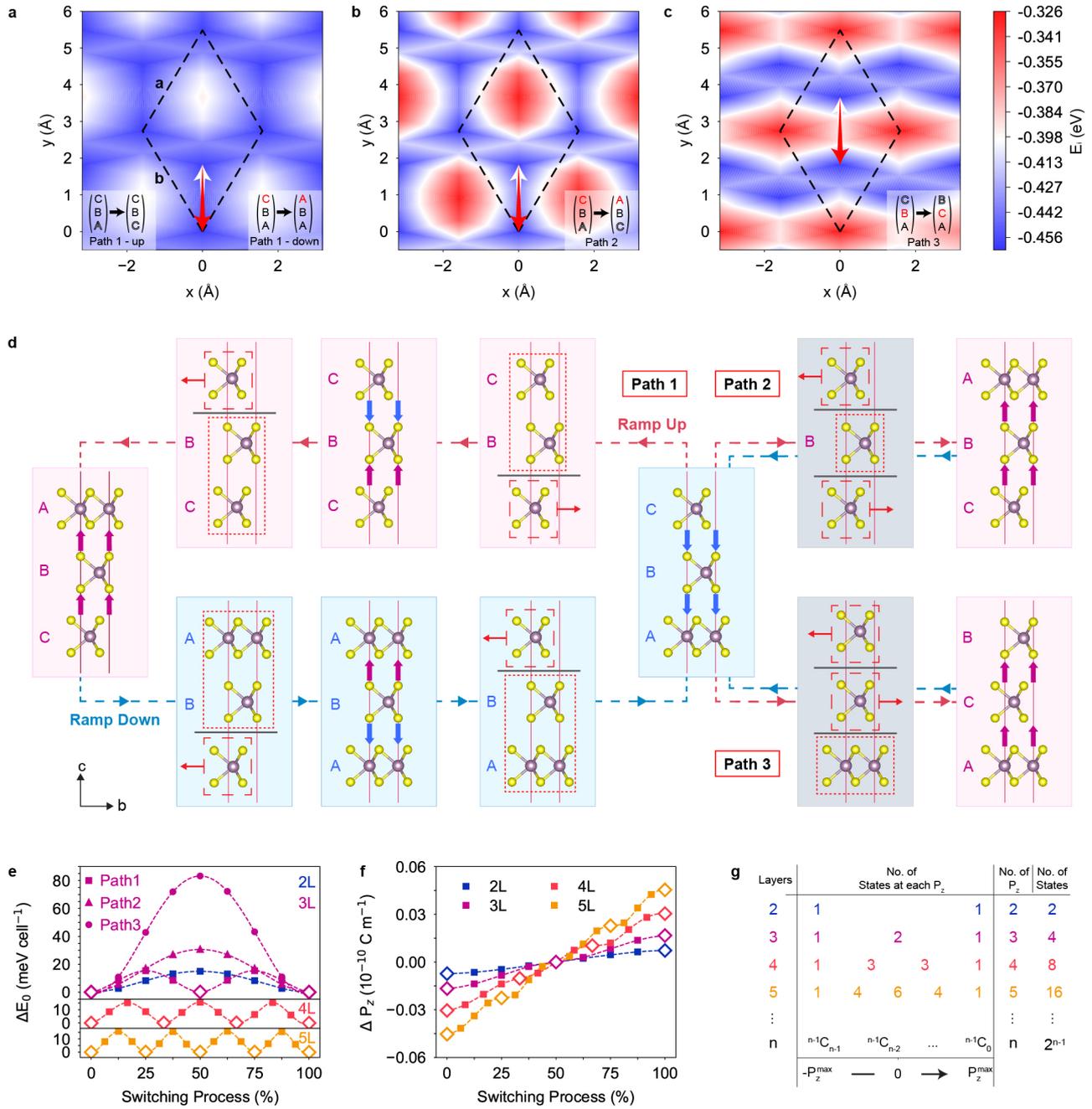

**Fig. 4. Theoretical analysis of ferroelectric switching in *n*-layer 3R MoS$_2$.** Potential energy surfaces experienced by the translating atomic layer(s) when **(a)** the other two layers are adjacent and stationery, **(b)** when they are not adjacent

but are translating simultaneously and **(c)** when they are adjacent and translating simultaneously. For ferroelectric switching pathways that match each of the above cases, arrows in (a-c) indicate the initial and final planar positions of the translating atomic layers' Mo atoms. The initial and final stacking orders of the translations are presented in the inset of each sub-figure. Translating atomic layers and their corresponding arrows in each sub-figure share the same colour. **(d)** Path 1, 2 and 3 of ferroelectric switching in trilayer 3R MoS$_2$. Atomic structures at 0%, 25%, 50%, 75% and 100% of the switching processes are shown. Red dashed (dotted) boxes indicate the moving (stationary) blocks. Gray horizontal lines indicate the sliding interfaces. The solid red horizontal arrows mark the displacement direction of the individual blocks. Atomic structures without the solid red horizontal arrows are stable structures. Purple and blue vertical arrows indicate the spontaneous polarization directions of the interface dipoles. High symmetry stacking positions (A, B or C) are labelled where applicable. **(e)** Total energy profiles of ferroelectric switching pathways in 3R MoS$_2$ of different thicknesses. Energy profiles of Paths 1, 2 and 3 are shown for trilayer (top), and energy profiles of paths for tetralayer (middle) and pentalayer (bottom) obtained from the generalized model are also shown. Energy of the initial state is taken as 0 eV cell$^{-1}$. **(f)** Variation of spontaneous polarization in $n$-layer 3R MoS$_2$ following the generalized model. Spontaneous polarizations of states at 50% of switching process are taken as zero. Negative $\Delta \mathbf{P}_z$ values denote downward polarization, and *vice versa*. In e and f, stable states are marked with hollow diamond markers. **(g)** Illustration detailing the relationship among the number of layers ($n$), the number of polarization states at each magnitude of spontaneous polarization ($N_P$), the total number of different spontaneous polarizations, and the total number of polarization states ($N$) in $n$-layer 3R MoS$_2$.